\input{aipcheck}

\documentclass[
    ,final            
  ]
  {aipproc}

\layoutstyle{8x11single}

\begin{document}

\title{Chiral-symmetry breaking and confinement in Minkowski~space}

\classification{11.30.Rd, 12.38.Lg, 12.39.Pn, 14.40.Be
                }
\keywords      {Chiral symmetry; quark model; Covariant Spectator Theory 
}

\author{
Elmar P. Biernat}{
address={Centro de F\'isica Te\'orica de Part\'iculas (CFTP), Instituto Superior T\'ecnico (IST), Universidade de Lisboa, 1049-001 Lisboa, Portugal}
}

\author{M. T. Pe\~na}{
address={Centro de F\'isica Te\'orica de Part\'iculas (CFTP), Instituto Superior T\'ecnico (IST), Universidade de Lisboa, 1049-001 Lisboa, Portugal}
  ,altaddress={Departamento de F\' isica, Instituto Superior T\'ecnico (IST), Universidadede Lisboa, 1049-001 Lisboa, Portugal}
}

\author{J. E. Ribeiro}{
  address={Centro de F\'isica das Intera\c c\~oes Fundamentais (CFIF), Instituto Superior T\'ecnico (IST), Universidade~de~Lisboa, 1049-001 Lisboa, Portugal} 
}
\author{Alfred Stadler}{
  address={Departamento de F\'isica, Universidade de \'Evora, 7000-671 \'Evora, Portugal
  }
  ,altaddress={Centro de F\'isica Te\'orica de Part\'iculas (CFTP), Instituto Superior T\'ecnico (IST), Universidade de Lisboa, 1049-001 Lisboa, Portugal}
}
\author{Franz Gross}{
  address={Thomas Jefferson National Accelerator Facility (JLab), Newport News, Virginia 23606, USA
  }
}

\begin{abstract}
We present a model for the quark-antiquark interaction formulated in Minkowski space using the Covariant Spectator 
Theory. The quark propagators are dressed with the same kernel that describes the interaction between different quarks. By applying the axial-vector Ward-Takahashi identity we show that our model satisfies the Adler-zero constraint imposed by chiral symmetry. For this model, our Minkowski-space results of the dressed quark mass function are compared to lattice QCD data obtained in Euclidean space. The mass function is then used in the calculation of the electromagnetic pion form factor in relativistic impulse approximation, and the results are presented and compared with the experimental data from JLab.   
\end{abstract}

\maketitle


\section{Introduction}

The pion remains an important system in hadronic physics to trace signatures of QCD in observables. It emerges non-perturbatively as the lightest quark-antiquark ($q\bar q$) bound state and it is at the same time identified with the Goldstone boson associated with spontaneous chiral-symmetry breaking (S$\chi$SB). The non-perturbative dynamics underlying such hadronic systems have been addressed in various modern approaches, such as QCD simulations on the lattice~\cite{Edwards,Guo}, light-front formulations of quantum field theory~\cite{Brodsky:1997de,Carbonell:1998rj,Sales:1999ec}, as well as models based on the Dyson-Schwinger--Bethe-Salpeter (DSBS) approach and the mass gap equation~\cite{Bars:1977ud,Amer:1983qa,LeYaouanc:1983it,Bicudo:1989sh,Bicudo:1989si,Bicudo:1989sj,Nefediev:2004by,Alkofer:2000wg,Maris:2003vk,Fischer:2006ub,Rojas:2013tza}, which have contributed to an understanding of a wide range of meson and baryon phenomena. We use the Covariant Spectator Theory (CST)~\cite{Gro69}, which is another modern approach that implements S$\chi$SB through the famous Nambu--Jona-Lasinio mechanism, similarly to DSBS. In contrast to the latter, whose four-dimensional integral equations are usually treated in Euclidean space, the CST equations can be solved directly in physical Minkowski space.

Previous CST quark models~\cite{Gross:1991te,Gross:1994he,Savkli:1999me} and improved versions of them that are currently being developed~\cite{PhysRevD.89.016005} employ an interaction kernel that includes linear confinement in a covariant-generalized form. Such kernels satisfy the relativistic version of the property that the nonrelativistic linear potential (in coordinate space) vanishes at the origin.\footnote{A useful method to treat the singularities of linear-confining interaction kernels has been recently proposed in Ref.~\cite{PhysRevD.90.096003}.} This implies that the confinement interaction \emph{decouples} from the CST-Dyson equation (CST-DE) for the scalar part of the dressed quark propagator, as well as from the CST pion equation in the chiral limit~\cite{Gross:1991pk}. 

At present, the precise Lorentz structure of the confining interaction is not known. Some approaches suggest that it has a large scalar component, and---although one lacks first-principle evidence for this---it is still quite important to study to what extent such confining forces can be made compliant with S$\chi$SB. In the previous CST models, a confining interaction was used that included non--chirally-symmetric spin structures, such as scalar terms, as they are not inconsistent with a massless solution of the pion equation in the chiral limit owing to the above discussed \lq\lq decoupling property.''

In the present work we follow a different strategy. We start from the most general Lorentz structure for the $q\bar q$ interaction and then determine the constraints imposed by S$\chi$SB, similarly to what has been done in Ref.~\cite{PhysRevD.47.1145}. It turns out that a CST model with scalar confinement, together with an equal-weighted pseudoscalar part satisfies the S$\chi$SB condition of the Adler consistency zero~\cite{Adler_PhysRev.137.B1022} in $\pi$-$\pi$ scattering in the chiral limit.
\section{Axial-vector Ward-Takahashi identity}
\begin{figure*}
\includegraphics[height=.18\textheight]{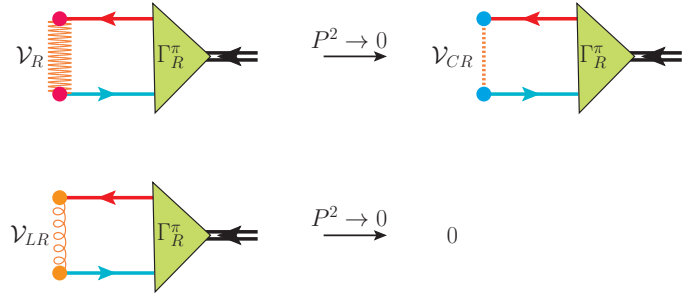}
\caption{(Color online) The decoupling of the linear-confining kernel from the pion CST-equation in the chiral limit. Each red or blue arrowed line denotes a dressed quark propagator. 
}\label{fig:pionChL}
\end{figure*}

Chiral symmetry and its breaking is expressed through the axial-vector Ward-Takahashi identity (AV-WTI), which is---within the Gross-Riska prescription~\cite{Gro87,Gro93,Gro96} of dealing with strong quark form factors---given by
\begin{eqnarray}
P_\mu \Gamma^{5\mu}_{R}(p',p)+2m_0 \Gamma^5_R(p',p)=\tilde S^{-1} (p')\gamma^5+\gamma^5\tilde S^{-1} (p)
\equiv\Gamma^{A}_R (p^\prime,p) \,,\label{eq:AVWTI}
\end{eqnarray}
 where $\Gamma^{5\mu}_R(p',p)$ and  $\Gamma^5_R(p',p)$ are the (reduced) dressed axial-vector and pseudoscalar vertex functions, respectively, $m_0$ is the bare quark mass, $p$ and $p'$ are the incoming and outgoing quark momenta, respectively, $P=p'-p$ is the momentum flowing into the vertex, $\tilde S (p)$ is the (damped) dressed quark propagator as introduced in Ref.~\cite{PhysRevD.90.096008}, and $\Gamma^{A}_R (p^\prime,p)$ is the (reduced) dressed \lq\lq axial vertex''---a convenient combination of the dressed axial-vector and the pseudoscalar vertices. $\Gamma^{A}_R (p^\prime,p)$ is the solution of an inhomogeneous CST Bethe-Salpeter equation (CST-BSE),
 \begin{eqnarray} 
\Gamma^{A}_R (p^\prime,p)=\gamma^{A}_R (p^\prime,p)+\mathrm i
\int_{k0} \mathcal V_R(p-k) 
\tilde S(k') \Gamma^{A}_R (k',k) \tilde S(k) \,, \qquad  \label{eq:CSTBSEGammaA}
\end{eqnarray}
 where $\gamma^{A}_R (p^\prime,p)$ is the (reduced) bare axial vertex, $\mathcal V_R(p-k)$ is the (reduced) covariant interaction kernel depending only on the four-momentum transfer $p-k=p'-k'$, and \lq\lq $k0$'' indicates the charge-conjugation invariant CST prescription for performing the $k_0$ contour integration~\cite{Savkli:1999me}. The most general structure of the linear-confining kernel, together with a vector--axial-vector remainder, is given by
 \begin{eqnarray}
 \mathcal V_R(p-k)= V_{LR}(p-k)\Big[\lambda_S ({\bf 1}\otimes {\bf 1})+\lambda_S (\gamma^5\otimes\gamma^5) +\lambda_V
( \gamma^\mu 
\otimes \gamma_{\mu})+ \lambda_A (\gamma^5\gamma^{\mu} \otimes \gamma^5\gamma_{\mu})+
\frac{\lambda_T}{2}(\sigma^{\mu\nu}\otimes\sigma_{\mu\nu})\Big]  \nonumber\\+V_{CR}(p-k)
\Big[\kappa_V(\gamma^\mu\otimes\gamma_\mu)+\kappa_A (\gamma^5\gamma^{\mu} \otimes \gamma^5\gamma_{\mu})\Big]
\,,\label{eq:kernel} 
\end{eqnarray}
where $V_{LR}$ and $V_{CR}$ are the momentum-dependent parts of the linear-confining and remaining kernels, respectively, with $V_{LR}$ satisfying 
\begin{eqnarray}
 \int \frac{\mathrm d^3 k}{E_k} V_{LR} (p\pm\hat k)=0\,,
 \label{eq:VLzero}
\end{eqnarray}
where $E_k=\sqrt{m^2+\vec k^2}$, $\hat k=(E_k,\vec k)$, and $m$ is the dressed quark mass. The corresponding weight parameters $\lambda_i$ and $\kappa_i$  [with $i=S$ (scalar), $P$ (pseudoscalar), $V$ (vector), $A$ (axial-vector), and $T$ (tensor)] are arbitrary constants, except that scalar and pseudoscalar parts in~(\ref{eq:kernel}) are \emph{equal-weighted}, i. e. $\lambda_S=\lambda_P$. For this kernel it has been shown~\cite{PhysRevD.90.096008} that the AV-WTI (\ref{eq:AVWTI}) together with the CST-BSE (\ref{eq:CSTBSEGammaA}) implies that $\tilde S (p)$ satisfies the CST-DSE,
\begin{eqnarray}
 \tilde S^{-1} (p)=\tilde S_0^{-1} (p)-\mathrm i\int_{k0} 
{\cal V}_R(p-k) \tilde S(k)\, ,
\label{eq:CST-DE2}
\end{eqnarray}
where $\tilde S_0$ is the (damped) bare quark propagator, which obeys the AV-WTI for an off-shell Ansatz of $\gamma^{A}_R (p^\prime,p)$ according to Gross and Riska. It turns out that $\gamma^{A}_R (p^\prime,p)$ vanishes in the chiral limit of vanishing bare quark mass, $m_0\rightarrow 0$, and vanishing vertex momentum, $P\rightarrow 0$.  In this limit, the CST-BSE~(\ref{eq:CSTBSEGammaA}) becomes homogeneous and identical to the zero-mass pion CST equation for the (reduced) pion vertex function in the chiral limit, $\Gamma^\pi_{R\chi}$, which implies the relation
\begin{eqnarray}
 \Gamma^\pi_{R\chi}(p,p)\propto\Gamma^{A}_{R\chi}(p,p)\,.
\label{eq:Gammachi}
\end{eqnarray}
Because of Eq.~(\ref{eq:VLzero}), only $\mathcal V_{CR}$ contributes to the chiral-limit pion equation and to the scalar part of the CST-DE~\eqref{eq:CST-DE2}, i.e. to dynamical quark mass generation. Therefore, the linear-confinement part $\mathcal V_{LR}$ that also includes scalar, pseudoscalar and tensor structures in our model, decouples from these equations. For the pion equation this is diagrammatically depicted in Fig.~\ref{fig:pionChL} and was proven in 
Ref.~\cite{PhysRevD.89.016005}. 
\section{$\pi$-$\pi$ scattering}
\begin{figure} 
\includegraphics[height=.4\textheight]{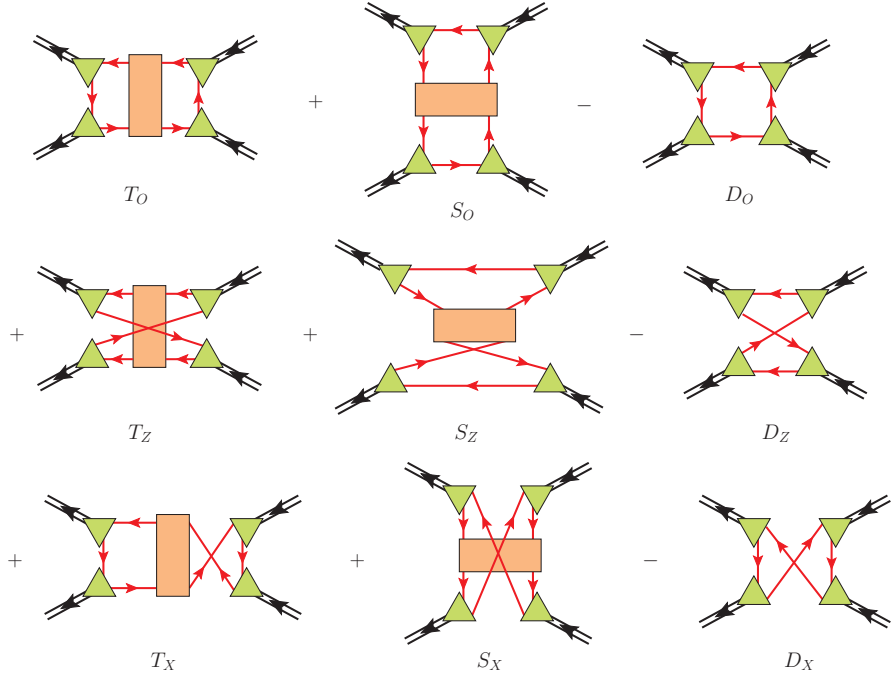}
   
  \caption{The contributions to $\pi$-$\pi$~scattering. The orange boxes denote the unamputated quark-quark scattering amplitudes.}
  \label{fig:pipiscattering}
\end{figure}

A stronger constraint for chiral symmetry than the previously discussed decoupling property of non--chirally-symmetric Lorentz structures is the Adler consistency zero of the $\pi$-$\pi$~scattering amplitude in the chiral limit. It has been realized earlier~\cite{PhysRevD.65.076008}, and also shown in CST~\cite{PhysRevD.90.096008}, that in order to obtain the Adler zero, it is essential to go beyond the impulse approximation in the scattering diagrams. In particular, for crucial cancellations in the amplitude to occur, it is unavoidable to include intermediate-state interactions to all orders through the complete quark-quark ladder sum. There are three types of contributions to the $\pi$-$\pi$~scattering amplitude, referred to as $O$, $Z$, and $X$ diagrams, as shown in Fig.~\ref{fig:pipiscattering}.
 It has been proven in CST that the sums of the three diagrams in each row vanish separately in the chiral limit~\cite{PhysRevD.90.096008}, i.e. 
 \begin{eqnarray}
  T_O+S_O-D_O\rightarrow0,\quad  T_Z+S_Z-D_Z\rightarrow0,\quad T_X+S_X-D_X\rightarrow0,\label{eq:chlOZXterms}
 \end{eqnarray}
 which constitutes the Adler zero. In the proof of (\ref{eq:chlOZXterms}) an additional ladder sum at one pion vertex has been inserted in the $T$ and the $S$ diagrams of Fig.~\ref{fig:pipiscattering} through the \lq\lq spectral decomposition''. Then, use of Eq.~(\ref{eq:Gammachi}) was made in the chiral limit to replace $\Gamma^\pi_{R\chi}$ by $\Gamma^{A}_{R\chi}$, allowing the application of the AV-WTI~\eqref{eq:AVWTI} between two ladder sums to reduce these diagrams to terms that cancel exactly the $D$ terms~\cite{PhysRevC.67.035201}. For our kernel \eqref{eq:kernel}, which includes scalar, pseudoscalar, and tensor structures that do not anticommute with $\gamma^5$, additional terms show up. Because of the decoupling of the linear-confining kernel from the zero-mass pion equation (Fig.~\ref{fig:pionChL}), these terms, however, vanish in the chiral limit (for details, see Ref.~\cite{PhysRevD.90.096008}).
\section{Dressed quark mass function}

In Ref.~\cite{PhysRevD.89.016005} we proposed a model for the $q\bar q$ interaction with a kernel of the form~\eqref{eq:kernel}, where $\mathcal V_{CR}$ is taken as a covariant momentum-space $\delta$-function with a pure vector structure ($\kappa_V=1$), and $\mathcal V_{LR}$ is a mixed scalar-vector linear-confining interaction, with $\lambda_S=2$ and $\lambda_V=1$ (with all other weight parameters set to zero). For this particular mixing the confining kernel does not contribute to the scalar part of Eq.~\eqref{eq:CST-DE2}, i.e. to the dressed quark mass. This leads to a rather simple dynamically-dressed quark mass function that is entirely determined by $\mathcal V_{CR}$. The same mass function is obtained from a similar kernel, with $\lambda_S=1$, $\lambda_P=1$, and $\kappa_V=1$ (with all other weight parameters set to zero), which now fully complies with S$\chi$SB.

\begin{figure}[h!] 
    \includegraphics[height=6cm]{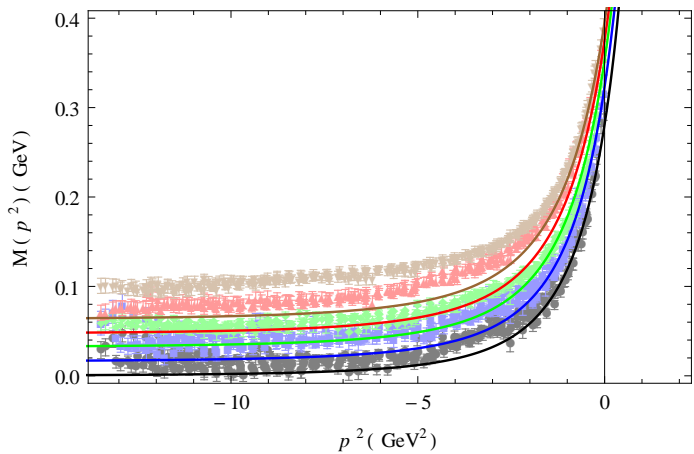} 
\caption{The dressed quark mass function $M(p^2)$ compared with lattice QCD data~\cite{Bowman:2005vx} for different bare quark masses  $m_0$. The five mass function curves and lattice data sets ($m_0=0$ data are extrapolated) from bottom to top correspond to $m_0=0$ (black, blobs),  $m_0=0.016$ GeV (blue, squares), $m_0=0.032$ GeV (green, diamonds), and $m_0=0.047$ GeV (red, triangles) and $m_0=0.063$ GeV (brown, inverted triangles).}\label{fig:1}
\end{figure}

Our mass function $M(p^2)$, obtained from solving the CST-DE~\eqref{eq:CST-DE2}, involves three free parameters: The dressed quark mass $m_\chi$ in the chiral limit, a mass parameter $M_g$ from the strong quark form factors, and the strength $C$ of the $\delta$-function kernel. Two of them, $m_\chi$ and $M_g$, are fixed in the chiral limit by a fit of the mass function to the lattice QCD data~\cite{Bowman:2005vx} extrapolated to $m_0=0$, which gives $m_\chi=0.308~\mathrm{GeV}$ and $M_g=1.734~\mathrm{GeV}$. The mass function for different values of $m_0$ is then found by solving the corresponding on-shell constraint $M(p^2=m^2)=m$, with $C$ as a function of $m_0$ chosen appropriately to fit the lattice data. The mass function result reads
\begin{eqnarray} 
 M(p^2)= \left(m_\chi+12 m_0\right)\,h^2(m^2)h^2(p^2)+m_0\,,\label{eq:1}
\end{eqnarray}
where
\begin{eqnarray}
 h(p^2)=\left(\frac{\Lambda_\chi^2-m_\chi^2}{\Lambda^2-p^2}\right)^2\, \label{eq:2}
\end{eqnarray}
are the strong quark form factors, with $\Lambda=m+M_g$ and $\Lambda_\chi=m_\chi+M_g$. The  Euclidean-space lattice data are compared with our Minkowski-space results at negative $p^2$, as shown in Figure~\ref{fig:1}.

\section{Electromagnetic pion form factor}
As a first test that our CST model gives sensible results we use the mass function~(\ref{eq:1}) (in the chiral limit) for the computation of the electromagnetic pion form factor, $F_\pi(Q^2,\mu)$, in relativistic impulse approximation (RIA)~\cite{Gross:1965zz,Arn77,Arn80,VO95}. The pion current in RIA---by keeping only the spectator quark propagator pole contribution of the triangle diagram---is given by
\begin{eqnarray}\label{eq:picurrentA}
 F_\pi(Q^2,\mu)(P_++P_-)^\mu
=\mathrm e\,\int \frac{\mathrm d^3 k}{(2\pi)^3}\frac{m}{E_k}\,\mathrm {tr}\Big[
\bar\Gamma_R^\pi (-\hat k,p_+) \tilde S(p_+)  j_R^\mu (p_+,p_-)\tilde S(p_-) \Gamma_R^\pi(p_-, -\hat k) \tilde\Lambda(-\hat k)\Big] \, 
\end{eqnarray}   
where $\mu$ is the pion mass, $P_\pm$ are the incoming and outgoing pion momenta, $p_\pm=P_\pm-\hat k$ are the off-shell quark momenta, $p_+-p_-=q$ with $Q^2=-q^2$ is the four-momentum transfer by the photon, $j^\mu_R$ is the reduced quark-photon vertex, and $\tilde\Lambda(\hat k)$ is the (appropriately-normalized) positive-energy Dirac projector. For the present simple calculation we adopt an approximated pion vertex function near the chiral limit, together with an Ansatz for the (reduced) dressed quark-photon vertex that is determined by the (vector) Ward-Takahashi identity,
\begin{eqnarray}
   q_\mu j^\mu_R(p_+,p_-)=\tilde S^{-1}(p_-)-\tilde S^{-1}(p_+)\,,
\end{eqnarray}
which ensures gauge invariance and, in particular, pion-current conservation. At a latter stage and for more realistic calculations we will use dynamically-calculated pion vertex functions and dressed quark-photon vertices obtained from solving the homogeneous pseudoscalar and inhomogeneous vector CST-BSE's, respectively. 

\begin{figure}
     \includegraphics[height=5.1cm]{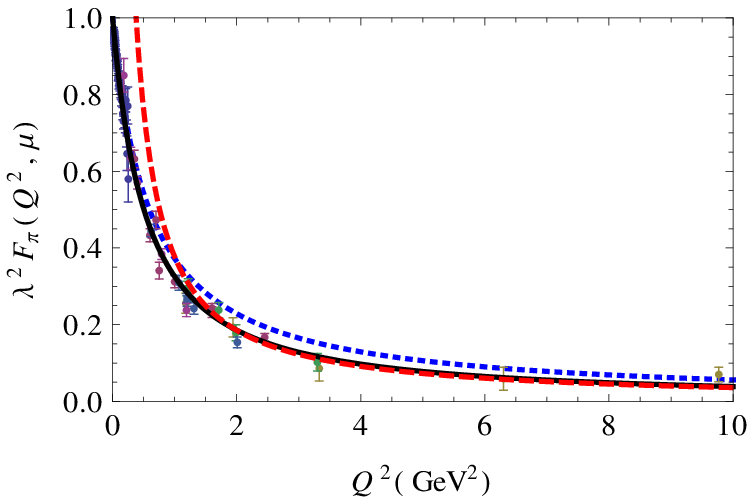} 
     \includegraphics[height=5cm]{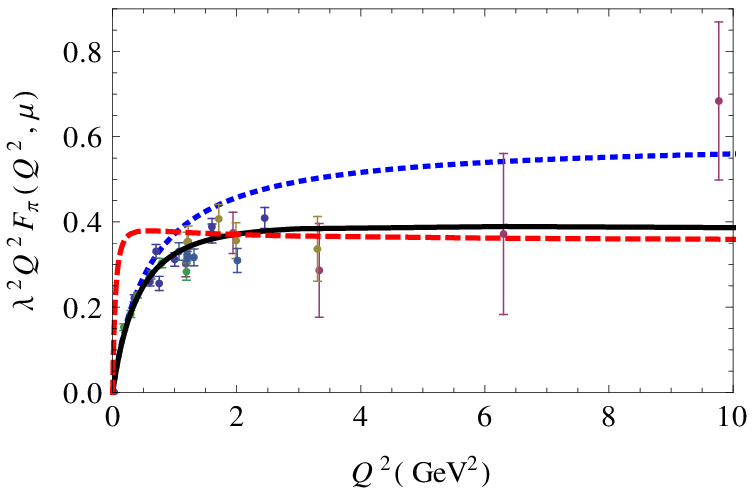} 
        \caption{The pion form factor compared to the JLab data~\cite{PhysRevC.78.045203}. 
      The left panel shows the pion form factor $F_\pi (Q^2,\mu=0.14)$ when scaled with $\lambda^2=(0.42/0.14)^2$ (red dashed line) to fit the pion form factor $F_\pi (Q^2,\mu=0.42)$ (black solid line) together with the $\rho$-pole contribution (blue dotted line). The right panel shows the same pion form factors but scaled with $Q^2$. } 
\label{fig:2}
\end{figure}

Remarkably, the result for $F_\pi(Q^2,\mu)$ in the present simple model is insensitive to the particular choice of the strong quark form factors $h$, but it depends on the pion mass $\mu$, in particular, at small $Q^2$. For sufficiently large $\mu$ (of the order of $m_\chi$ and larger), one expects the RIA to be a good approximation for the full triangle diagram of the pion form factor, not only at high but also at low $Q^2$~\cite{PhysRevD.89.016006}. In particular, the value $\mu$=0.42 GeV gives the best fit to the experimental data~\cite{PhysRevC.78.045203} over the full (spacelike) range of $Q^2$. At large $Q^2$ we find an interesting scaling behavior between the pion form factors calculated with different pion masses:
\begin{eqnarray}
   F_\pi(Q^2,\lambda \mu)\stackrel{Q^2\gg\mu^2}{\simeq}\lambda^{2} F_\pi(Q^2,\mu)\,, \label{eq:scaling}
\end{eqnarray} 
 where $\lambda$ is a scaling parameter. In Fig.~\ref{fig:2} the large-$Q^2$ tail of the form factor calculated with the physical value $\mu=0.14$ GeV is scaled to fit the one calculated with $\mu=0.42$ GeV.

It should be emphasized that this is a very simple model for the pion form factor that, for instance, does not include explicit contributions from the $\rho$ meson. Such contributions should be important in the timelike region according to vector meson dominance, and thus we expect this model to fail in this region. Nevertheless, in the spacelike region our result exhibits the correct monopole behavior at large $Q^2$ and is in good agreement with the experimental data. This concludes the first qualitative study of our model, which shows that it is able to give sensible results for both the dressed quark mass function and the spacelike pion form factor, at the same time. It is clear that for a more quantitative study of the pion structure that also extends to timelike $q^2$, the solutions of the CST-BSE's are needed, with a dynamically-dressed quark current that includes the $\rho$-pole contribution. 

As a long-term goal we plan to calculate transition form factors, to be used---together with the dressed quark currents---in the calculation of hadronic contributions to light-by-light scattering, such as pseudoscalar-meson pole and dressed-quark loop contributions. They are essential ingredients in precision calculations of the anomalous magnetic moment of the muon that could reveal new physics beyond the Standard Model.


\begin{theacknowledgments}
This work received financial support from Funda\c c\~ao para a Ci\^encia e a 
Tecnologia (FCT) under Grants No.~PTDC/FIS/113940/2009 and No. CFTP-FCT (PEst-OE/FIS/U/0777/2013). The research leading to these results has received funding from the European Community's Seventh Framework Programme FP7/2007-2013 under Grant Agreement No.\ 283286. This work was also partially supported by Jefferson Science Associates, LLC, under U.S. DOE Contract No. DE-AC05-
06OR23177. 
\end{theacknowledgments}



\bibliographystyle{aipproc}   

\bibliography{PapersDB}

\IfFileExists{\jobname.bbl}{}
 {\typeout{}
  \typeout{******************************************}
  \typeout{** Please run "bibtex \jobname" to obtain}
  \typeout{** the bibliography and then re-run LaTeX}
  \typeout{** twice to fix the references!}
  \typeout{******************************************}
  \typeout{}
 }

\end{document}